# A Deep Representation Empowered Distant Supervision Paradigm for Clinical Information Extraction

Yanshan Wang, PhD[1], Sunghwan Sohn, PhD[1], Sijia Liu, MS[1], Feichen Shen, PhD[1], Liwei Wang, PhD[1], Elizabeth J. Atkinson, MS[1], Shreyasee Amin, MD, MPH[2,3], Hongfang Liu, PhD[1]

[1]Division of Biomedical Statistics and Informatics, Department of Health Sciences Research, Mayo Clinic, Rochester, MN
[2]Division of Rheumatology, Department of Medicine, Mayo Clinic, Rochester, MN
[3]Division of Epidemiology, Department of Health Sciences Research, Mayo Clinic, Rochester, MN

**Abstract**

*Objective: To automatically create large labeled training datasets and reduce the efforts of feature engineering for training accurate machine learning models for clinical information extraction.*

*Materials and Methods: We propose a distant supervision paradigm empowered by deep representation for extracting information from clinical text. In this paradigm, the rule-based NLP algorithms are utilized to generate weak labels and create large training datasets automatically. Additionally, we use pre-trained word embeddings as deep representation to eliminate the need of task-specific feature engineering for machine learning. We evaluated the effectiveness of the proposed paradigm on two clinical information extraction tasks: smoking status extraction and proximal femur (hip) fracture extraction. We tested three prevalent machine learning models, namely, Convolutional Neural Networks (CNN), Support Vector Machine (SVM), and Random Forrest (RF).*

*Results: The results indicate that CNN is the best fit to the proposed distant supervision paradigm. It outperforms the rule-based NLP algorithms given large datasets by capturing additional extraction patterns. We also verified the advantage of word embedding feature representation in the paradigm over term frequency-inverse document frequency (tf-idf) and topic modeling representations.*

*Discussion: In the clinical domain, the limited amount of labeled data is always a bottleneck for applying machine learning. Additionally, the performance of machine learning approaches highly depends on task-specific feature engineering. The proposed paradigm could alleviate those problems by leveraging rule-based NLP algorithms to automatically assign weak labels and eliminating the need of task-specific feature engineering using word embedding feature representation.*

**Introduction**

The initialization of the Health Information Technology for Economic and Clinical Health Act (HITECH Act) in 2009 has fostered the rapid adoption of Electronic Health Record (EHR) systems at US hospitals and clinics. Large amounts of detailed longitudinal patient information, including lab tests, medications, disease status, and treatment outcomes, have been accumulated and are available electronically and become valuable data sources for clinical and translational research [1-3]. A well-known challenge faced when using EHR data for research is that large amounts of detailed patient information is embedded in clinical text (e.g., clinical notes and progress reports). Information extraction, one of the popular Natural language processing (NLP) technologies, can unlock information embedded in clinical text by automatically extracting desired information (e.g. cancer stage information [4-6], disease characteristics [7-9] and pathological conditions [10]) from the text. Many successful studies applying information extraction techniques have been reported, ranging from phenotyping [11, 12], detecting adverse events [13], improving healthcare quality [14, 15] to facilitating genomics research such as gene-disease association analysis [16, 17] and pharmacogenomic studies[18, 19].

Clinical information extraction applications can be developed using either symbolic techniques or statistical machine learning [20]. Applications built based on symbolic techniques involve handcrafted expert rules, such as regular expressions and logic rules. It has been shown effective in the clinical domain due to the clinical sublanguage characteristics [21]. However, rule-based applications can be expensive and cumbersome to develop, requiring the collaboration between NLP experts and healthcare professionals, and the resultant applications may not be portable. Meanwhile, machine learning approaches have been shown efficient and effective [22, 23] for clinical information extraction tasks, for example, disorder normalization [24], gout flares extraction [25], cancer identification [26], or thromboembolic diseases identification [27]. Despite their impressive improvements, large amounts of manually labeled training data are a crucial building block and a key enabler for a successful machine learning model. However, such large labeled training data is not always readily available and usually expensive to create due to human annotation. This problem becomes more significant in the clinical domain, mainly due to i) the lack of publicly available clinical corpora because of privacy concerns, and ii) the annotation of clinical text requiring medical knowledge. Therefore, popular methods for creating labeled training data for clinical information extraction tasks, such as crowdsourcing, are not applicable. The distant supervision strategy has been utilized to quickly create large training data in the literature. It uses existing resources or a set of heuristics [28] to generate weakly labeled training data, which has been applied in common NLP tasks including relation extraction [29, 30], knowledge base completion [31], sentiment analysis [32], and information retrieval [28]. In addition to the need of large labeled training data, machine learning requires feature engineering where each training instance needs to be transformed into feature vector representation. Recently, the deep representation learning has become popular due to its capability of representing the raw data (such as the pixel values of an image or words in a textual document) in a high level representation or feature vector [33]. In NLP, word embeddings are one of the most successful deep learning technologies with the ability to capture high-level semantic and syntactic properties of words [34-36]. Word embeddings have been utilized in various clinical NLP applications, such as clinical abbreviation disambiguation [37], named entity recognition [38], and information retrieval [39].

In this study, we propose a distant supervision paradigm empowered by deep representation for extracting information from clinical text. The distant supervision paradigm utilizes the rule-based NLP algorithms to generate weak labels and create large training data automatically. The deep representation using pre-trained word embeddings makes it effortless for feature engineering. In the following, we provide background and related work of the study, followed by a detailed description of our methods. We then conducted experiments using two case studies at Mayo Clinic to illustrate the effectiveness of the proposed paradigm: smoking status extraction and proximal femur (hip) fracture extraction. We tested three prevalent machine learning models, namely, Convolutional Neural Networks (CNN), Support Vector Machine (SVM), and Random Forrest (RF), and verified the advantage of word embedding features in the proposed paradigm. Moreover, we showed the impact of the training data size on the performance of machine learning methods.

**Background and Related Work**

As an applied domain, information extraction from clinical text has been dominated by rule-based NLP approaches. According to a recent literature review [20], more than 60% of the clinical information extraction studies published after 2009 used only rule-based NLP systems. The reasons are due to its interpretability to clinicians and its ability to incorporate domain knowledge from knowledge bases or experts, which is essential for clinical applications. Successful use cases of clinical NLP include identification of diseases, such as Kawasaki disease [40], detection of pathological causes of obesity [41], and ascertainment of patient asthma status from clinical notes [42-45]. In this study, we utilize a list of keywords and regular expression patterns in an NLP tool, MedTagger [46], to extract information from clinical documents. MedTagger constitutes the Apache UIMA pipeline and NLP techniques to facilitate information extraction from free-text EHRs, which are tailored to specific clinical information extraction tasks.

Word embeddings have been widely adopted in clinical information extraction applications as they could encode deep semantic representations of words [38, 47, 48]. Henriksson et al. [47] leverage word embeddings to identify adverse drug events from clinical notes. Their experiments show that employing word embeddings could improve the predictive performance of machine learning methods. Both Tang et al. [48] and Wu et al.'s [38] studies show that word embedding features outperform other features for clinical named entity recognition. In addition, word embeddings help improve the relation extraction, such as relations between medical problems and treatments, relations between medical problems and tests, and relations between medical problems and medical problems, in clinical notes [49]. However, to the best of our knowledge, there is no study in the literature utilizing word embeddings as features to empower the distant supervision approach.

Distant supervision is a simple and adaptable approach for programmatic creation of labeled training sets. It is proposed primarily for relation extraction from text, wherein a known relation from a knowledge base (e.g. Freebase) is likely to express that relation in an input corpus [29, 30]. Later, distant supervision has been widely applied in other common NLP tasks including knowledge-base completion [31], sentiment analysis [32], and information retrieval [28]. In the biomedical domain, distant supervision has been used to augment machine learning based classifiers to identify drug-drug interactions or medical terms from biomedical literature [50-53]. In the clinical domain, Wallace et al. [54] proposed a distant supervision approach to better exploit a weakly labeled corpus to extract sentences of population/problem, intervention, comparator, and outcome from clinical trial reports. However, to the best of our knowledge, the distant supervision has not been utilized for clinical information extraction tasks. Particularly, no study utilizes rule-based NLP for generating weakly labeled training data for machine learning methods empowered by word embedding features. Our hypothesis is that the deep representation of word embeddings might enable machine learning methods to learn extra rules from weak labeled training data and outperform the rule-based NLP systems used to generate the weak labels since it could find semantically similar words in embedding space while these words may not be included in the NLP rules.

**Methods**

*Distant Supervision Paradigm*

We now describe the deep representation empowered distant supervision paradigm for clinical information extraction from clinical text. In this paradigm, information extraction is defined as a text classification task, aiming to assign the appropriate categories to the text. Figure 1 illustrates the schema of the proposed distant supervision paradigm.

A simple rule-based NLP algorithm is firstly developed based on knowledge bases or expert knowledge. Then the rule-based NLP algorithm is applied on non-labeled clinical text to extract the desired information. For each instance, a weak label is assigned to indicate whether such information is extracted. By doing so, one can create a large set of weakly labeled training data quickly.

The pre-trained word embeddings are used to map each instance into a deep semantic vector representation. Machine learning methods are then trained using the vector representations as input and the corresponding weak labels as learning objectives. Eventually we utilize the trained machine learning model to extract information from unseen clinical text.

We theoretically show in the Supplement that training machine learning models with the true labels can be approximated using the weak labels generated by the rule-based NLP and machine learning models trained from weak labels can achieve similar training performance compared to the models trained from true labels. Thus, the proposed distant supervision paradigm addresses the problem of lacking large labeled training data for machine learning models without hurting the performance for clinical information extraction.

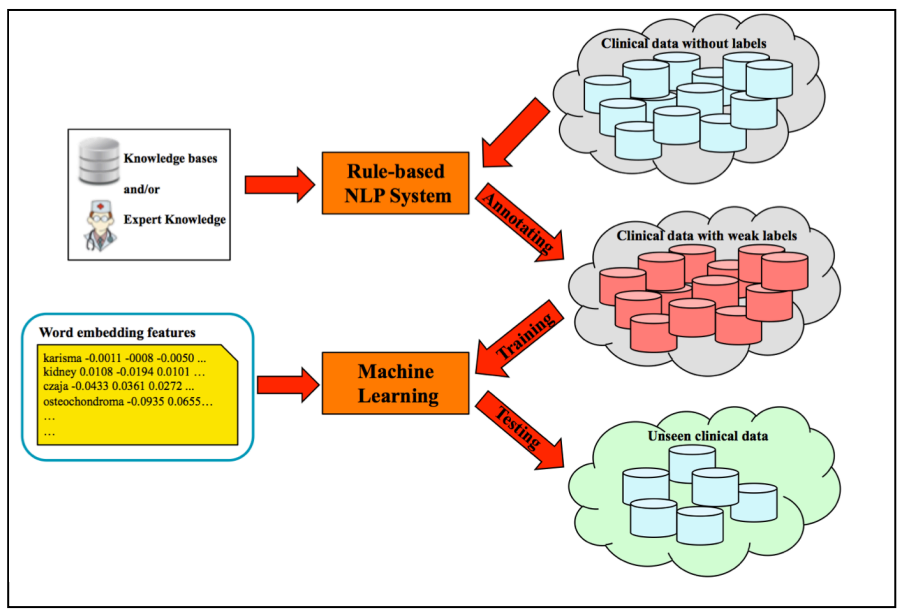

**Figure 1.** The schema of the deep representation empowered distant supervision paradigm.

*Machine Learning Methods*

Although any machine learning method can be applied in the proposed paradigm, we would like to investigate which model fits better in the paradigm. In this study, we tested three prevalent machine learning methods in the literature, namely Support Vector Machine (SVM), Random Forest (RF), and Convolutional Neural Networks (CNN). SVM is a supervised learning that has been widely used for classification [55]. RF is an ensemble of classification trees, where each tree contributes with a single vote for the assignment of the most frequent class to the input data [56]. Compared to SVM, RF usually has high classification accuracy and ability to model complex interactions among input variables. CNN is a feed-forward artificial neural network with layers formed by a convolution operation followed by a pooling operation [57]. In our experiment, we used a simple CNN model consisting of embedding layer, convolution layer and fully-connected layer with a softmax function, as shown in Figure 2.

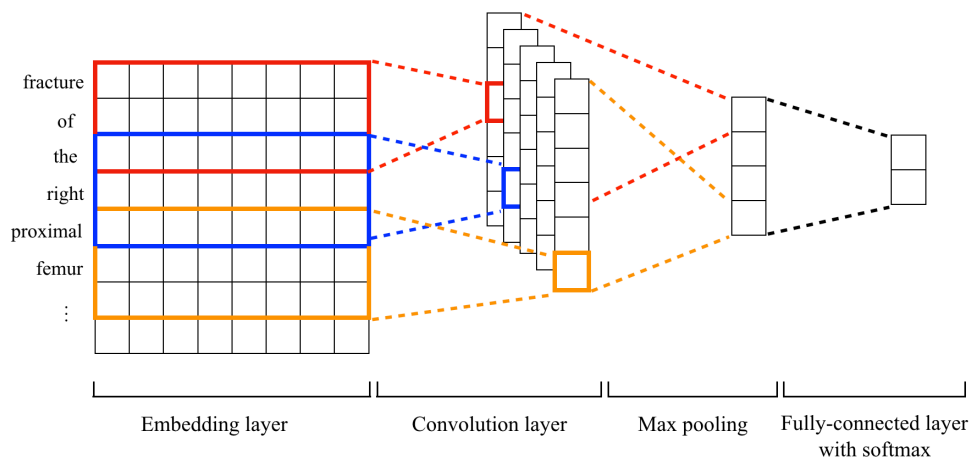

**Figure 2.** Architecture overview of the CNN model.

*Word Embeddings*

The word embeddings used in our experiments were trained by word2vec [58] on a large corpus consisting of the textual clinical notes for a cohort of 113k patients receiving their primary care at Mayo Clinic, spanning a period of 15 years from 1998 to 2013 [59]. We set the dimension of word embeddings to 100 since word embeddings with this dimension size can represent medical word semantics better than those with other dimension sizes [59]. For CNN, the pre-trained word embeddings are directly utilized to map words into vectors in the embedding layers. In order to obtain the feature of each instance for SVM and RF, we calculate the mean of the summation of word embeddings of words in the instance. Specifically, given an instance $d = \{w_1, w_2, .., w_M\}$ where $w_i, i = 1,2, ..., M$ is the $i$th word and $M$ is the total number of words in this instance, the feature vector $\mathbf{x}$ of instance $d$ is defined by:

$$\mathbf{x} = \frac{1}{M} \sum_{i}^{M} \mathbf{x}_i,$$

where $\mathbf{x}_i$ is the embedding vector for word $w_i$ from the word embeddings.

**Materials**

We evaluated the effectiveness of the proposed paradigm on two clinical information extraction tasks: smoking status extraction and proximal femur (hip) fracture extraction. This study was approved by the institutional review board (IRB) for human subject research.

*Case Study 1: Smoking Status Extraction*

We first examined the proposed distant supervision paradigm on a smoking status extraction task at Mayo Clinic with the aim of identifying the smoking status in a clinical note, i.e., smoker (including current smoker and past smoker) or non-smoker. We curated a corpus of 32,336 instances by using the "social and behavior history" section from the clinical notes in the Mayo Clinic EHR system [60, 61]. To evaluate the performance, we randomly sampled 475 of them to create a test dataset with the gold standard labels manually annotated by an expert with medical background. For the remaining 31,861 clinical notes, we developed a simple rule-based NLP algorithm, as shown in Table 1, to extract smoking status instances. Note that the smoking status was non-smoker if no information was extracted from a clinical note. By doing so, we created a large weakly labeled training dataset for machine learning models.

**Table 1.** Keywords of the NLP algorithm for the extraction of smoking status.

| Smoker | smokes?, smoked, smoking, smokers?, tobaccos?, cigarettes?, cigs?, pipes?, nicotine, cigars?, tob |
|---|---|
| Non-Smoker | (no\|non\|not\|never\|negative)\W*(smoker\|smoking\|smoked\|tobacco), nonsmoker, denies\W*smoking, (tobacco\|smoke\|smoking\|nicotine)\W*(never\|no), doesn\'t smoke, 0\|zero smokers? |

*Case Study 2: Proximal Femur (Hip) Fracture Extraction*

In the second experiment, we evaluated the distant supervision paradigm on a proximal femur (hip) fracture extraction task at Mayo Clinic. Among fractures, proximal femur (hip) fractures are of particular clinical interest as they are most often related to bone fragility from osteoporosis, and are associated with significant mortality and morbidity in addition to high health care costs [62]. In this task, a set of 22,969 radiology reports (including general radiography reports, computed tomography reports, magnetic resonance imaging reports, nuclear medicine radiology reports, mammography reports, ultrasonography reports, neuroradiology reports, etc.) from 6033 Mayo Clinic patients were used to determine whether a proximal femur (hip) fracture could be identified using radiology reports [63, 64]. The subjects were aged 18 years of age or older, were residents of Olmsted County, and had experienced at least one fracture at some site during 2009-2011. Similar to the previous experiment, we first randomly sampled 498 radiology reports as testing data and asked a medical expert with multiple years of experience abstracting fractures to assign a gold standard to each radiology report.

Table 2 shows the rule-based NLP algorithm for this proximal femur (hip) fracture extraction task. The rules were developed and refined through verification with physicians and supplemented with historical rules developed by the Osteoporosis Research Program at Mayo Clinic to aid the nurse abstractors in proximal femur (hip) fracture extraction. In this NLP algorithm, the fracture modifiers must appear in the context of keywords within a sentence. We ran this NLP algorithm on the training dataset and obtained a weak label for each document, which we subsequently trained using the machine learning methods. Finally, we tested the performance on the testing dataset using the gold standard obtained similarly as the smoking status task.

**Table 2.** Keywords of the NLP algorithm for the extraction of proximal femur (hip) fracture.

| Keywords | cervical\|femoral head\|neck, (trans)?cervical, (sub)?capital, intracapsular, trans( \|-)?epiphyseal, base of neck, basilar femoral neck, cervicotrochanteric, (greater\|lesser) trochanter, (inter\|per\|intra) trochanteric |
|---|---|
| Fracture Modifiers | (micro-?)?fracture(s\|d)?, (epi\|meta)physis, separation, fxs?, broken, cracked, displace(d)?, fragment |

*Evaluation Metrics*

Precision, recall, and F1 score, defined below, were used as metrics to evaluate the performance.

$$\text{Precision} = \frac{\text{TruePositive}}{\text{TruePositive} + \text{FalsePositive}}$$

$$\text{Recall} = \frac{\text{TruePositive}}{\text{TruePositive} + \text{FalseNegative}}$$

$$\text{F1 score} = 2 \cdot \frac{\text{Precision} \cdot \text{Recall}}{\text{Precision} + \text{Recall}}$$

# Results

## Results of the Distant Supervision Paradigm

Figure 3 shows the results of the proposed distant supervision paradigm compared with the rule-based NLP algorithms, where DS-SVM, DS-RF and DS-CNN represent the approaches using SVM, RF, and CNN in the distant supervision paradigm, respectively. It shows that DS-CNN achieves the best performance among the tested machine learning methods, and slightly outperforms the rule-based NLP algorithms for both tasks with statistical significance ($p < 0.01$). The results may reflect that CNN was able to capture hidden patterns missed by a rule-based NLP algorithm from the weakly labeled training data. DS-SVM is inferior to DS-CNN for the smoking status extraction, but comparable to DS-CNN for the proximal femur (hip) fracture extraction. The performance of DS-RF is worse than DS-CNN for both extraction tasks. DS-RF performs better than DS-SVM for the smoking status extraction while worse for the proximal femur (hip) fracture extraction. The results from both experiments show that the DS-CNN is the best fit in the proposed distant supervision paradigm and could outperform the rule-based NLP algorithms.

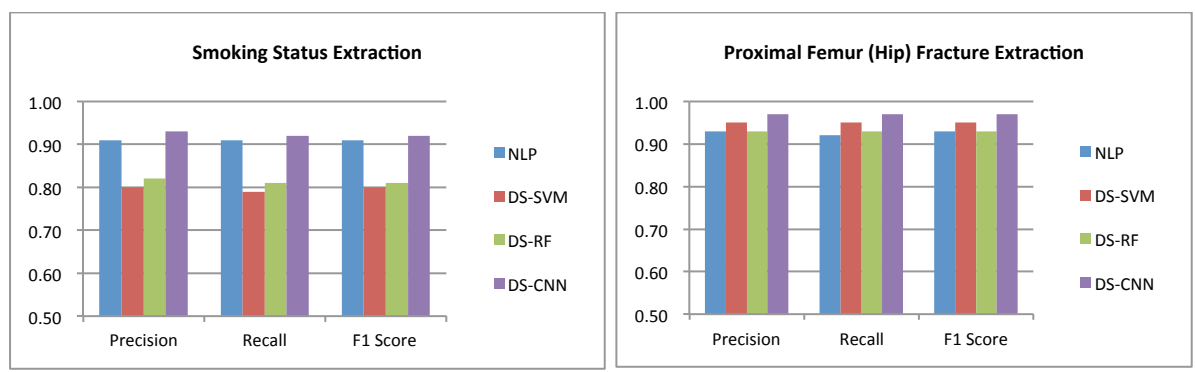

**Figure 3.** Comparison results of the distant supervision paradigm. NLP represents the rule-based NLP algorithms. DS-SVM, DS-RF and DS-CNN represent the approaches using SVM, RF, and CNN in the distant supervision paradigm, respectively.

## Impact of the Word Embedding Features

To verify the advantage of work embeddings in the proposed distant supervision paradigm, we compared the performance of word embeddings in SVM and RF with two other popular document representations: term frequency-inverse document frequency (tf-idf) and topic modeling. Since the layer of word embeddings is a component of the CNN model, this comparison is not conducted on CNN. The tf-idf document representation is a common term weighting scheme in information retrieval, which has been also found effective for document classification [65-67]. It represents a document using a vector with dimension as the vocabulary size of the corpus and elements corresponding to the tf-idf weight of each word $w$ in the document $d$, denoted by

$$tf - idf(d, w) = tf(w) \cdot idf(d, w),$$

where

$$tf(w) = \frac{\text{\# term } w \text{ in the doc } d}{\text{total \# terms in the doc } d},$$

$$idf(d, w) = \log \frac{\text{total \# docs in the corpus}}{\text{\# docs with the term } w \text{ in the corpus}}.$$

To derive document representation using topic modeling, we employed Latent Dirichlet Allocation (LDA) [68] and set the number of topics as 100. The document representation was then derived similarly as word embeddings where $x_i$ is now the word-topic mixture distribution of word $w_i$.

The results are shown in Figure 4. The models using word embedding features perform better than those using tf-idf and topic modeling features with statistical significance ($p < 0.01$). The reason might be that word embedding features could alleviate the feature sparsity issues compared to tf-idf features [69], and represent better semantics compared to topic modeling features [70]. The models using topic modeling features are better than those using tf-idf features since topic modeling features are also semantic representations of words. Since topic modeling requires prior distributions that are always difficult to define for a given corpus [71], it is usually inferior to word embeddings. This experiment verifies the advantage of word embeddings used as features for machine learning models in the proposed distant supervision paradigm.

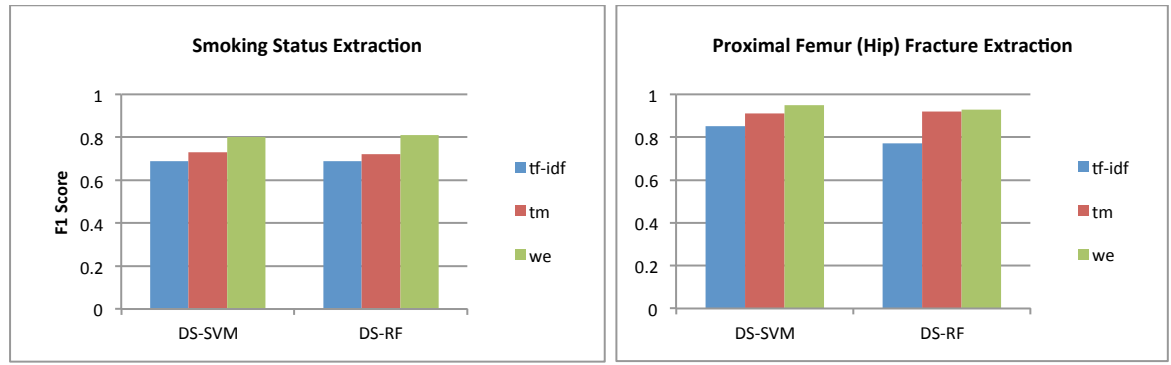

**Figure 4.** Comparison of using different numbers of documents in the fracture task. **tf-idf** represents the term frequency-inverse document frequency feature, **tm** represents the topic modeling feature, **we** represents the word embedding feature.

*Impact of the Data size*

One question we are interested is: do we really need the entire dataset of 31,861 clinical notes for training in the smoking status extraction task or that of 22,471 radiology reports for training in the proximal femur (hip) fracture extraction task? In order to answer this question, we tested the proposed paradigm by using different sizes of training data, namely 1,000, 2,500, 5,000, 10,000, and 20,000. Note that these training data were randomly sampled from the entire dataset. Figure 5 depicts the F1 score curves of machine learning methods and the rule-based NLP algorithms for the smoking status extraction and proximal femur (hip) fracture extraction tasks. When the data size is 1,000, DS-SVM and DS-RF outperform DS-CNN in both tasks. When the data size increases to 5000, the performance of DS-CNN increases dramatically and becomes better than DS-SVM and DS-RF in both tasks. As the data size becomes 10,000, DS-CNN does not have much performance gain compared to the data size of 5,000, but it outperforms the rule-based NLP algorithm for both tasks. When the data size is 20,000, the performance of DS-CNN is the same as that when the data size is 10,000. The performance curves of DS-CNN clearly show that this deep learning method is more sensitive to the data size, while the rule-based NLP algorithms and the conventional machine learning methods are more resistant to the data size. We can also see that 5,000 training data might be sufficient for DS-CNN to learn most extraction patterns for a single concept.

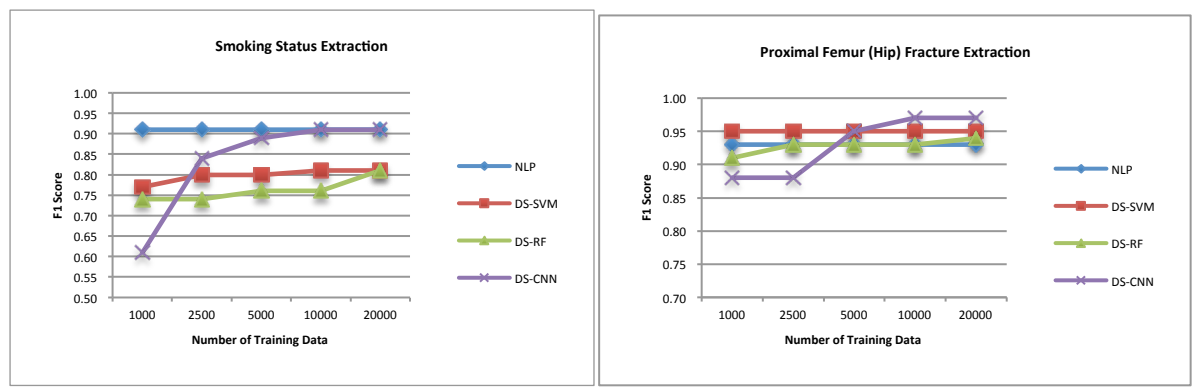

**Figure 5.** Comparison of using different sizes of training dataset. NLP represents the rule-based NLP algorithms. DS-SVM, DS-RF and DS-CNN represent the approaches using SVM, RF, and CNN in the distant supervision paradigm, respectively.

*Extra Patterns Learned by CNN*

In this section, we compare the distant supervision paradigm using DS-CNN and the rule-based NLP algorithm, and demonstrate additional extraction patterns could be captured by DS-CNN compared with the rule-based NLP.

In the smoking status extraction task, the information extracted by DS-CNN and that by the rule-based NLP algorithm is different for 7 out of 475 testing cases. Among these 7 cases, DS-CNN identified the current smoking status for 5 cases (71.4%). Table 3 lists three typical cases where the results of DS-CNN and the rule-based NLP algorithm are different. In Case 1, the rule-based NLP algorithm failed due to the misspelling word "tobaco" in the clinical note, which was not considered in the rules. DS-CNN was able to address this issue since it used word embedding features that could represent the misspelled word as being closely related to the correct form in the semantic space based on its context [59]. The rule-based NLP incorrectly extracted non-smoker information from Case 2 due to the pattern "no smoking". However, the whole statement "no smoking after age XXX" indicates a past smoker. This semantic meaning could be captured by DS-CNN. DS-CNN failed in Case 3 where the rule-based NLP algorithm correctly captured the correct smoking status due to the rules inspired by human experience. Many physicians write clinical notes following certain structures, which result in semi-structured clinical note, such as "Tobacco current use: No never used any" in Case 3. Since we were aware of this based on experts' experience, the rules in the NLP algorithm could handle it properly. However, machine learning methods might focus on the pattern "Tobacco current use" and thus extracted the incorrect smoking status.

The proximal femur (hip) fracture identified by DS-CNN differs from that by the rule-based NLP algorithm for 6 out of 498 testing cases. Among these 6 cases, DS-CNN correctly identified the hip fracture status for 5 cases (83.3%). Similar to the smoking status extraction task, we list a few typical cases in Table 3 for the hip fracture extraction task. In Case 1, the rule-based NLP algorithm failed to extract fracture information since the NLP failed to match the pattern of the keyword "fx" and "right femoral neck", which occurred across sentences. DS-CNN has no such issue since it does not require a sentence detection algorithm. Case 2 is likely a fracture in proximal femur where a percutaneous pin has been placed. Similar to Case 1 in the smoking status extraction task, the rule-based NLP algorithm failed in Case 2 due to the missing fracture keyword "fracture" in the report. Unlike the rule-based NLP algorithm that entirely relies on the rules, DS-CNN could correctly extract the fracture information since the representations of these keywords in the embedding space are semantically similar to "fracture". Case 3 is not describing a proximal femur fracture based on the context. However, the rule-based NLP algorithm matched the rules in the sentence "Fx…cervical" and ignored the context in the middle. In contrast, DS-CNN could take the context into account for calculating the note representation using word embedding features and accurately determined this is not a proximal femur fracture. The rule-based NLP algorithm correctly extracted fracture information for Cases 4 whereas DS-CNN failed. The reason might be that the proximal femur fracture signal in this case is too weak for DS-CNN as the note only mentioned the fracture in the indication.

**Table 3.** Cases from the experimental extraction tasks where the results of DS-CNN and the rule-based NLP algorithm are different. We use Y to indicate the correct extraction result and X otherwise.

| Task | Case # | Text Snippets | Gold Standard | Rule-based NLP | DS-CNN |
|---|---|---|---|---|---|
| Smoking Status Extraction | 1 | *…She is a taxi driver and she has never used tobaco products…* | Non-smoker | X | Y |
| | 2 | *…No smoking after age XXX…* | Smoker | X | Y |
| | 3 | *…Tobacco current use: No never used any...* | Non-smoker | Y | X |
| Proximal Femur (Hip) Fracture Extraction | 1 | ...Indications: femur fx...Cannulated screw fixation of the right femoral neck... | Proximal Femur fracture | X | Y |
| | 2 | ... Pin fixation across the proximal left femoral neck... | Proximal Femur fracture | X | Y |
| | 3 | Exam: Sp Cerv*2vw Flex/Ext only Indications: Fx Vertebra Cervical Closed... | Non-Proximal Femur fracture | X | Y |
| | 4 | Exam: R Major Jnt Asp and/or Inj Indications: R hip inj/marc/steroid; fx femur neck nos closed, pain hip... | Proximal Femur fracture | Y | X |

Table 4 lists some keywords in our extraction tasks and the selected semantically similar words found by the deep representation method. First, we can observe that the deep representation could capture similar words regardless of morphological change. For example, "cigar", "cigarettes", and "cigars" are similar words to "cigarette", "fx" is similar to "fracture". More interestingly, it could find misspelled words similar to the correct forms, such as "nicotene" to "nicotine", "cervial" to "cervical", and "intratrochanteric" and "introchanteric" to "intertrochanteric". Second, the deep representation method could find semantically similar words. For example, "cigarette" is semantically similar to "tobacco"; "intramedullary", "intermedullary", "nailing", and "pinning" are related to surgical fixation of the hip fracture; "transtrochanteric", "pertrochanteric", "basicervical", and "intertroch" are similar to the keyword "intertrochanteric" in terms of a description of the location of the proximal femur. Keywords of either different morphologies or semantics may not be easily identified by humans when developing a rule-based NLP algorithm.

**Table 4.** Keywords for the extraction tasks and the corresponding semantically similar words found by the deep representation method.

| Task | Keyword | Selected semantically similar words |
|---|---|---|
| Smoking Status Extraction | smoke | secondhand, thirdhand, pipes, nutcrackers, cigs |
| | tobacco | cigarettes, cigarette, cigar, cigars, tobaco |
| | cigarette | cigar, hookah, tobacco, cigarettes, cigars |
| | nicotine | nicotene, nicoderm, nictoine |
| Proximal Femur (Hip) Fracture Extraction | fx | fracture, comminuted, pinning, displaced, fractures |
| | intertrochanteric | intramedullary, nailing, pinning, intratrochanteric, introchanteric, transtrochanteric, pertrochanteric, basicervical, intertroch |
| | greater trochanter | trochanters, troch, trochanteric |

**Discussion**

In the clinical domain, the limited amount of labeled data is always a bottleneck for applying machine learning and large scale annotation is hampered by legal and privacy constraints on sensitive health records [72]. The proposed paradigm could alleviate this problem by leveraging the rule-based NLP algorithm to automatically assign weak labels. Along with automatic generated word embedding features, we demonstrate that our approach can achieve high performance on clinical information extraction tasks. Compared to the conventional NLP and machine learning methods, our approach needs much less human efforts in terms of knowledge engineering and training data annotation. Since there are publicly available NLP algorithms [20] and pre-trained word embeddings [73], the proposed paradigm may be an easier way for non-experts to use machine learning methods for extracting clinical information from clinical text in healthcare institutions.

Our study shows that deep neural networks are robust to massive label noise and a sufficiently large training data is important for effectively training deep neural networks, consistent with a recent study in the common machine learning domain[74]. A shortcoming of deep neural networks is that they lack the interpretability. This drawback is explicitly shown in our error analysis. Rules can be added or modified in rule-based NLP algorithms but it is difficult to revise trained deep neural network models.

The deep word embedding representation could identify semantically similar words, which enabled deep neural networks to learn more hidden patterns. As a matter of fact, these semantically similar words could also augment the rules in the NLP algorithm to improve identifying positives. However, the deep representation also found noisy words that are irrelevant to the specific information extraction task. In the future work, we would like to study how to leverage word embeddings to add effective keywords into the rule-based NLP algorithm.

Evaluation of the portability of distant supervision paradigm is also an interesting topic. Since clinical practice and workflow vary across institutions, the performance of NLP systems varies across institutions and sources of data [75]. An NLP system performing well in one institution might need to redesign rules according to the sublanguage characteristic in the institutional EHR system, which requires lots of efforts. However, machine learning models may not need extra modification when switching from one institution to another institution since they learn rules automatically, which may significantly reduce implementation time and expenses. Therefore, the portability of the proposed paradigm across different institutions is an important research topic and subject to a future study.

A few questions regarding the theory of distant supervision paradigm remains open. For example, it is not clear how simple an NLP algorithm (i.e., how many rules) is sufficient for machine learning methods and what accuracy an NLP algorithm should have (i.e., how small should $\epsilon$ be) to generate useful weak labels. Although we show that 5,000 training data was sufficient for DS-CNN for concept extraction tasks (i.e., smoking status or proximal femur (hip) fracture), it still remains an open question how much data are adequate for training machine learning models for complex extraction tasks.

**Limitations**

There are some limitations of this study. First, our experiments are simple tasks that aim to extract single concepts. Evaluation of the proposed paradigm for extracting multiple concepts in a complex task (such as multiclass text classification) is subject to a future study. Second, we did not use shared task datasets, such as the i2b2 (Informatics for Integrating Biology to the Bedside) 2006 smoking status extraction dataset [76], to evaluate the proposed approach. The reason is that the size of shared task datasets is usually very small (e.g., only a set of 889 discharge summaries are provided by i2b2 shared task), which may not be feasible to the proposed distant supervision paradigm.

**Conclusions**

In this paper, we proposed a deep representation empowered distant supervision paradigm for information extraction from free-text EHRs. In this paradigm, we utilize the output of rule-based NLP algorithms as weak labels to curate a large set of training data, and leverage pre-training word embedding features to represent the training data for machine learning methods. Although the training data is weakly labeled, we theoretically show that machine learning models trained from these weak labels can achieve similar training performance to that trained from true labels. We validated the effectiveness of the proposed paradigm using two clinical natural language processing tasks: a smoking status extraction task and a fracture extraction task, and tested three prevalent machine learning models, i.e., SVM, RF, and CNN. The results from both experiments show that the CNN is the best fit in the proposed paradigm that could outperform the rule-based NLP algorithms. We also verified the advantage of word embedding features in the proposed paradigm over two other widely adopted features: tf-idf and topic modeling. Moreover, we found that the CNN captures additional extraction patterns compared with the rule-based NLP but is more sensitive to the size of training data.

**Acknowledgements**

This work was supported by NIA P01AG04875, NIGMS R01GM102282, NCATS U01TR002062, and NLM R01LM11934 and made possible by the Rochester Epidemiology Project (NIA R01AG034676) and the U.S. Public Health Service.
**References**

[1] St. Sauver JL, Grossardt BR, Yawn BP, Melton III LJ, Rocca WA. Use of a medical records linkage system to enumerate a dynamic population over time: the Rochester epidemiology project. American journal of epidemiology. 2011;173:1059-68.
[2] Dean BB, Lam J, Natoli JL, Butler Q, Aguilar D, Nordyke RJ. Use of electronic medical records for health outcomes research: A literature review. Medical Care Research and Review. 2009;66:611-38.
[3] Jensen PB, Jensen LJ, Brunak S. Mining electronic health records: towards better research applications and clinical care. Nature Reviews Genetics. 2012;13:395-405.
[4] Cheng LT, Zheng J, Savova GK, Erickson BJ. Discerning tumor status from unstructured MRI reports—completeness of information in existing reports and utility of automated natural language processing. Journal of digital imaging. 2010;23:119-32.
[5] Nguyen AN, Lawley MJ, Hansen DP, Bowman RV, Clarke BE, Duhig EE, et al. Symbolic rule-based classification of lung cancer stages from free-text pathology reports. Journal of the American Medical Informatics Association. 2010;17:440-5.
[6] Warner JL, Levy MA, Neuss MN, Warner JL, Levy MA, Neuss MN. ReCAP: Feasibility and accuracy of extracting cancer stage information from narrative electronic health record data. Journal of oncology practice. 2015;12:157-8.
[7] Zhu H, Ni Y, Cai P, Qiu Z, Cao F. Automatic extracting of patient-related attributes: disease, age, gender and race. Studies in health technology and informatics. 2012;180:589-93.
[8] Shen F, Wang L, Liu H. Using Human Phenotype Ontology for Phenotypic Analysis of Clinical Notes. Studies in health technology and informatics. 2017;245:1285-.
[9] Shen F, Wang L, Liu H. Phenotypic Analysis of Clinical Narratives Using Human Phenotype Ontology. Studies in health technology and informatics. 2017;245:581-5.
[10] Séverac F, Sauleau EA, Meyer N, Lefèvre H, Nisand G, Jay N. Non-redundant association rules between diseases and medications: an automated method for knowledge base construction. BMC medical informatics and decision making. 2015;15:29.
[11] Liao KP, Cai T, Savova GK, Murphy SN, Karlson EW, Ananthakrishnan AN, et al. Development of phenotype algorithms using electronic medical records and incorporating natural language processing. bmj. 2015;350:h1885.
[12] Shen F, Liu S, Wang Y, Wang L, Afzal N, Liu H. Leveraging Collaborative Filtering to Accelerate Rare Disease Diagnosis.  American Medical Informatics Association2017.
[13] Rochefort C, Verma A, Eguale T, Buckeridge D. O-037: Surveillance of adverse events in elderly patients: A study on the accuracy of applying natural language processing techniques to electronic health record data. European Geriatric Medicine. 2015;6:S15.
[14] St-Maurice J, Kuo MH, Gooch P. A proof of concept for assessing emergency room use with primary care data and natural language processing. Methods Inf Med. 2013;52:33-42.
[15] Hsu W, Han SX, Arnold CW, Bui AA, Enzmann DR. A data-driven approach for quality assessment of radiologic interpretations. Journal of the American Medical Informatics Association. 2015;23:e152-e6.
[16] McCarty CA, Chisholm RL, Chute CG, Kullo IJ, Jarvik GP, Larson EB, et al. The eMERGE Network: a consortium of biorepositories linked to electronic medical records data for conducting genomic studies. BMC medical genomics. 2011;4:13.
[17] Zhang Y, Shen F, Mojarad MR, Li D, Liu S, Tao C, et al. Systematic identification of latent disease-gene associations from PubMed articles. PloS one. 2018;13:e0191568.
[18] Denny JC, Ritchie MD, Basford MA, Pulley JM, Bastarache L, Brown-Gentry K, et al. PheWAS: demonstrating the feasibility of a phenome-wide scan to discover gene–disease associations. Bioinformatics. 2010;26:1205.
[19] Xu H, Jiang M, Oetjens M, Bowton EA, Ramirez AH, Jeff JM, et al. Facilitating pharmacogenetic studies using electronic health records and natural-language processing: a case study of warfarin. Journal of the American Medical Informatics Association : JAMIA. 2011;18:387-91.